\newcommand{\be}{\begin{equation}}
\newcommand{\ee}{\end{equation}}
\newcommand{\bea}{\begin{eqnarray}}
\newcommand{\eea}{\end{eqnarray}}
\def\pslash{{\cal P}{\hbox{\kern-6pt $\slash$}}}

\long\def\comment#1{}

\documentstyle[prl,aps,floats,twocolumn]{revtex}

\input{psfig.tex}
\begin{document}
\draft
\twocolumn[\hsize\textwidth\columnwidth\hsize\csname @twocolumnfalse\endcsname
\title{The Second Peak: The Dark-Energy Density and the Cosmic
    Microwave Background}
\author{Marc Kamionkowski and Ari Buchalter}
\address{Mail Code 130-33, California Institute of Technology,
Pasadena, CA 91125}
\date{January 2000}
\maketitle

\begin{abstract}
Supernova evidence for a negative-pressure dark energy (e.g.,
cosmological constant or quintessence) that contributes a
fraction $\Omega_\Lambda\simeq0.7$ of closure density
has been bolstered by the discrepancy between the total density, $\Omega_{\rm
tot}\simeq1$, suggested by the location of the first peak in the
cosmic microwave background (CMB) power spectrum and the
nonrelativistic-matter density $\Omega_m\simeq0.3$ obtained from 
dynamical measurements.  Here we show that the impending
identification of the location of the {\it second} peak in the
CMB power spectrum will provide an immediate and independent
probe of the dark-energy density.  As an aside, we
show how the measured height of the first peak probably already
points toward a low matter density and places upper limits to
the reionization optical depth and gravitational-wave amplitude.
\end{abstract}

\pacs{PACS number(s):
98.80.-k,95.35.+d,98.70.Vc,98.65.Dx,98.80.Cq \hfill
CALT-68-2253}
]

A ``cosmic-concordance'' model now seems to be falling into
place \cite{CC}.  The central and most intriguing feature of this
model is a negative-pressure dark energy
(e.g., cosmological constant or quintessence) that contributes a
fraction $\Omega_\Lambda\simeq0.7$ of closure density.
Supernova evidence for this dark energy \cite{supernovasearches}
has been bolstered by
the discrepancy between the total density, $\Omega_{\rm
tot}\equiv\Omega_m+\Omega_\Lambda\simeq1$, suggested by the
location of the first peak in the cosmic microwave background
(CMB) power spectrum \cite{flat,boomerang} and the
nonrelativistic-matter density, $\Omega_m\simeq0.3$, obtained from 
dynamical measurements.  

This dark energy has implications of the utmost
importance not only for cosmology, but for fundamental physics
as well.  It can be viewed equivalently/alternatively as a
correction to general relativity or as some new exotic form of
matter.  It would have significant implications for the
evolution of large-scale structure in the Universe, for particle 
theory, and possibly for quantum gravity.  Theorists have
expanded the realm of possibilities for this dark energy from a
simple cosmological constant to quintessence, a variable
cosmological constant driven by the rolling of some new scalar
field \cite{quint,quintcmb}.  Given the extraordinary ramifications, it
is crucial to test for a nonzero dark-energy density as
thoroughly as possible.  There are already several promising
possibilities; e.g.,  statistics of gravitational-lens systems
\cite{QSO}, the Alcock-Paczy\'nski test \cite{AP}, and
cross-correlation of the CMB with some tracer of the density
at lower redshifts \cite{XRB}.

The purpose of this article is to show that impending
measurements of the location of the {\it second} peak in the CMB
power spectrum will provide an additional and independent probe
of the dark-energy density.  We argue that the location of the
second peak depends primarily on the matter density and on the
Hubble constant ($h$ in units of 100 km~sec$^{-1}$~Mpc$^{-1}$),
and plot contours of the second-peak location
in the $\Omega_m$-$h$ parameter space.  If the Hubble constant
is fixed by independent observations (e.g., from the Hubble
Space Telescope [HST]), then the
second-peak location determines the matter density, or
equivalently, the dark-energy density.  As an aside, we also
illustrate how recent measurements of the height of the first
peak may already be pointing to a low value of $\Omega_m$.

\begin{figure}[t]
\centerline{\psfig{file=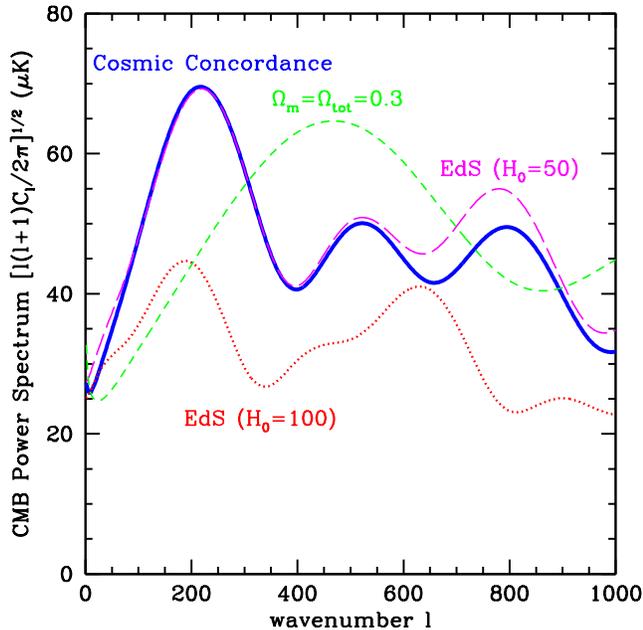,width=3.5in}}
\caption{A few illustrative CMB power spectra.  The heavy solid
     (blue) curve is the cosmic-concordance model with
     $\Omega_m=0.3$, $\Omega_\Lambda=0.7$, $h=0.7$, and
     $\Omega_b h^2=0.019$ (and a touch of reionization).
     Current data (not shown) indicate a first peak at
     $l\simeq200$ with an amplitude around 70 $\mu$K.  The
     short-dash (green) curve shows that an open model
     with $\Omega_m=\Omega_{\rm tot}=0.3$ produces a peak far
     to the right of that observed.  The dotted (red) curve, an
     Einstein-de-Sitter model with a Hubble constant $h=1$,
     illustrates that in models with both a high density and a
     high Hubble constant, the first peak is too low to match
     that observed.  It also demonstrates how the second peak
     gets absorbed into the third peak.  The long-dash (magenta) 
     curve shows that an Einstein-de-Sitter Universe can produce 
     a power spectrum that agrees with that of the
     cosmic-concordance model up through the first two peaks,
     but only with a Hubble constant well below the HST value.}
\vskip -0.5cm
\label{fig:Clsplot}
\end{figure}

The aim of CMB mapping experiments is to measure the temperature 
$T(\hat n)$ as a function of position $\hat n$ on the
sky \cite{KamKos99}.
The temperature can then be expanded in spherical harmonics,
$a_{lm} = \int\, Y_{lm}(\hat n) \, T(\hat n)$,
and rotationally invariant multipole moments (the ``power
spectrum''), $C_l = \sum_m |a_{lm}|^2/(2l+1)$,
can be constructed.  Given the values of several cosmological
parameters, predictions for the power spectrum can be made;
Fig. \ref{fig:Clsplot} shows a few models.  The peak structure
is due to oscillations in the primordial plasma \cite{SZ}.

The location in $l$ of the first peak depends strongly on
$\Omega_{\rm tot}$ and only very weakly on the values of
other cosmological parameters, and so it provides a robust
indicator of the geometry of the Universe \cite{kss}.  A compilation of data
from a number of recent experiments indicates a peak near
$l\sim200$, and data from the test flight of BOOMERANG clearly
shows a peak at this location.  We thus assume that, as argued
by Dodelson and Knox \cite{flat}, the verdict is in: the
Universe is flat.\footnote{Strictly speaking, such a peak
location could be fit in an open or closed Universe with
combinations of very strange values for other cosmological
parameters (e.g., \cite{ZalSelSpe97}).  We realize this as a
mathematical possibility, but a physical improbability.}

If the geometry is fixed, the location of the second peak in the
CMB power spectrum depends primarily (though not entirely) on
the expansion rate of the Universe at the epoch of recombination
\cite{husug}, and this depends on the nonrelativistic-matter density
and the Hubble constant.  In principle, variations in several other
parameters can change the precise location of the second peak.
However, the second-peak location shifts very little as each of
these uncertain parameters is allowed to vary within its
acceptable range.  For example, if measurements of the deuterium
abundance fix $\Omega_b h^2=0.019\pm0.001$ \cite{tytler}, as
Tytler asserts, then allowable shifts in the
baryon-to-photon ratio produce negligible shifts in the
location of the second peak.  To be safe, we show results below
for the more conservative range, $0.015<\Omega_b h^2<0.023$,
advocated by Olive, Steigman, and Walker \cite{osw}.  The
spectral index $n$ of primordial density perturbations changes
the amplitudes of the peaks, but allowable variations in $n$
($\pm0.3$ \cite{COBE}) lead to even smaller uncertainties in 
the second-peak location than those from uncertainty in the
baryon density.  Moreover, a more recent
analysis that includes constraints from degree-scale CMB
anisotropies and large-scale structure finds that the
allowed range for $n$ is much tighter---within 5\% of unity
\cite{bondjaffe}.
Reionization may reduce
the amplitudes of all the peaks, but it will not strongly affect
their locations, and the same is true of gravitational  waves.
Plausible neutrino masses would have a negligible
effect on the peak locations \cite{neutrinos}.  Higher-order
effects, such as weak gravitational lensing \cite{wl}, the
Rees-Sciama effect \cite{rs}, or unsubtracted foregrounds would
primarily affect the heights or shapes of the peaks but leave
their locations intact.  The second-peak location is similarly
insensitive to whether the dark energy is a cosmological
constant or quintessence \cite{quintcmb}.  We also expect
magnetic fields to have no more than a small effect on the peak
location \cite{magnetic}.

\begin{figure}[t]
\centerline{\psfig{file=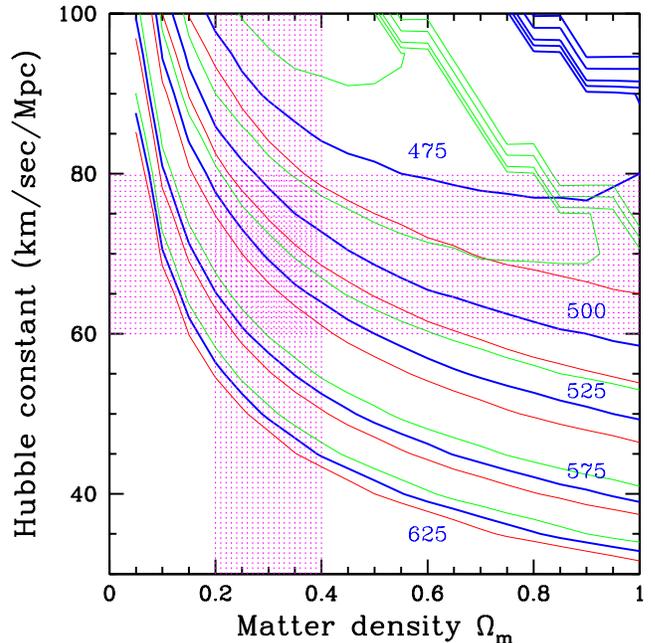,width=3.5in}}
\caption{Contours of the multipole moment $l_2$ at which the
     second peak in the CMB power spectrum occurs.  The heavy
     solid curves (blue) show contours of $l_2=$475, 500, 525,
     575, and 625 for the central value, $\Omega_b h^2 =0.019$,
     of the baryon-to-photon ratio.  The lower lighter (red)
     curves and the upper lighter (green) curves show the same
     for $\Omega_b h^2=0.015$ and $\Omega_b h^2=0.023$,
     respectively.  The horizontal and vertical shaded regions
     are those allowed, respectively, by HST measurements of the
     Hubble constant and by supernova results.  The
     cosmic-concordance model lies at the intersection of these
     two.}
\vskip -0.5cm
\label{fig:secondpeak}
\end{figure}

Fig. \ref{fig:secondpeak} shows contours of $l_2$, the location
of the second peak, in the two-dimensional
parameter space ($\Omega_m$,$h$) in which it varies most
strongly.  Results are shown for the allowable range of
$\Omega_b h^2$.  Had we included contours for $n=0.8$ and
$n=1.2$, they would have fallen very well within the range
spanned by the allowed values of the baryon-to-photon ratio.

The location of the second
peak picks out a specific contour in the $\Omega_m$-$h$ plane.
When combined with the range, $0.6<h<0.8$ \cite{HST}, allowed by
HST, determination of the second-peak location will provide a
constraint to the matter density.  For example, if the second
peak turns out to be located at $l_2\gtrsim625$, then it will
suggest $\Omega_m\lesssim0.2$ (for the entire allowable range
for $\Omega_b h^2$).  A value $l_2\simeq550$ will
allow $0.1\lesssim\Omega_m\lesssim0.4$ (likewise, for the
allowed range of $\Omega_bh^2$).  If it turns out that
$l_2\simeq525$, then a broader range of larger values,
$0.2\lesssim\Omega_m \lesssim0.6$, will be allowed.  Smaller
values of $l_2$ allow larger values of $\Omega_m$, and they
are also less constraining.  If $\Omega_m\lesssim0.4$, as
suggested by supernova data \cite{supernovasearches}, then the
second peak must be at $l_2>475$.

The contours in Fig. \ref{fig:secondpeak} show that $l_2$ jumps
to very large values for large $\Omega_m$ and large $h$ (the
upper right-hand corner).  In this region of parameter space,
the amplitude of the second peak actually becomes so small that
the second peak disappears, and the {\it de facto} second peak
is what would have otherwise been the third peak.  To
illustrate, the long-dash (magenta) curve in Fig. \ref{fig:Clsplot} shows the
$C_l$ for $\Omega_m=0.8$, $h=0.9$, and $\Omega_b h^2=0.015$.
The region of the $\Omega_m$-$h$ parameter space in which this
confusion between the second and third peaks arises conflicts
with the age of the Universe, the shape of the power spectrum,
and as discussed below, the amplitude of the first peak.  We
therefore dwell no further on this possibility.

There are some caveats we should make.  The allowed range of
$\Omega_m$ for any given value of $l_2$ can be broadened if a
larger range of values for the Hubble constant are allowed.  So,
for example, if the second peak is found to be at $l_2=525$, it
will be possible that $\Omega_m=1$, but only if the Hubble
constant is $h=0.5$, considerably lower than the HST value.
(This could be tested further with the
third peak, as indicated in Fig. \ref{fig:Clsplot}.)
A smaller baryon density would shift the second peak to larger
values of $l$ and thus allow slightly larger values of
$\Omega_m$ for fixed $l_2$.  However, such small values of the
baryon density would conflict not only with Tytler's results,
but would be additionally discordant with baryon abundances in
x-ray clusters.  Some combination of other effects (e.g., the
optical depth, primordial spectrum, neutrino masses,
recombination history, etc.) could move the peak, but such a
conspiracy seems unlikely.  Thus, the weakest link in the
relation between $l_2$ and $\Omega_m$ is
probably uncertainty in the Hubble constant, as indicated in
Fig. \ref{fig:secondpeak}.  Even if independent measurements of
the Hubble constant are discarded, the location of the second
peak will provide a useful constraint to the $\Omega_m$-$h$
parameter space.

\begin{figure}[t]
\centerline{\psfig{file=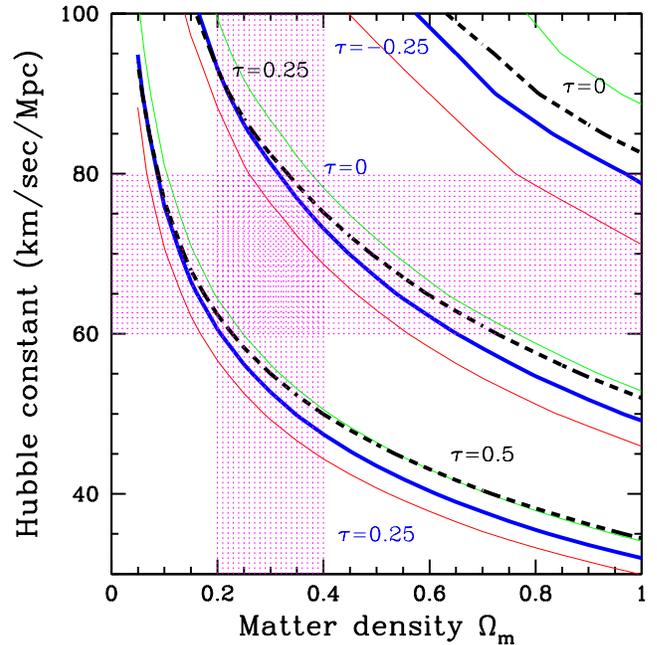,width=3.5in}}
\caption{Contours of the value of the optical depth
     $\tau$ to the surface of last scatter inferred by comparing 
     the predicted amplitude of the first peak with the measured 
     value of $\simeq70$ $\mu$K.  The heavy solid (blue) curves show
     contours of $\tau=0.25$, 0, and $-0.25$ for the central
     value, $\Omega_b h^2 =0.019$ and for an $n=1$ primordial
     spectrum of density perturbations.  The lower lighter (red)
     curves and the upper lighter (green) curves show the same
     for $\Omega_b h^2=0.015$ and $\Omega_b h^2=0.023$,
     respectively.  Since $\tau<0$ is impossible, the
     high-$\Omega_m$--high-$h$ regions in which the Figure
     indicates that $\tau<0$ are ruled out.  The dashed
     curves show contours of $\tau=0$, 0.25, and 0.5 for the
     central value of the
     baryon density, but for a primordial spectral index
     $n=1.2$.  Again, the horizontal and vertical shaded regions 
     are those allowed, respectively, by HST measurements of the
     Hubble constant and by supernova results.}
\vskip -0.5cm
\label{fig:tauplot}
\end{figure}

The main focus here is on the location of the second peak.
However, it is easy and important to see that the observed
height of the first peak already points toward a low density if
primordial perturbations have a flat scale-invariant (i.e.,
$n=1$) spectrum.  It is natural to expect that at least some
small fraction $\tau$ of CMB photons re-scattered from reionized 
electrons after the nominal surface of last scatter at redshift
$z\simeq1100$.  If so, then the amplitude of the peaks in the
power spectrum will be suppressed by a factor
$e^{-\tau}$.
Fig. \ref{fig:tauplot} shows
contours of the optical depth $\tau$ inferred by comparing the
predicted height of the first peak with the measured value of
$\simeq70$ $\mu$K
for the allowable range of $\Omega_b h^2$, and for a
flat (i.e., $n=1$) primordial spectrum and for an $n=1.2$
primordial spectrum.  Since $\tau<0$ is impossible, those
regions of parameter space in which $\tau<0$ is inferred are
ruled out.  If primordial perturbations have a flat
spectrum, then the amplitude of the first peak thus rules out a
considerable portion of the high-$\Omega_m$--high-$h$ parameter
space.  Moreover, notice that a shift of 0.2 in $n$ is roughly
equivalent to a shift of about 0.25 in $\tau$.  Thus,
constraints to the $\Omega_m$-$h$ parameter space from the height
of the first peak can be relaxed if the spectral index $n$ is
raised a bit, while models with $n\simeq0.8$ are likely
inconsistent as they would require a negative optical depth over 
virtually the entire plausible range of $\Omega_m$ and $h$.  The
constraints may also be relaxed if the actual 
amplitude is a bit different than the value, 70 $\mu$K, used
here, as may be allowed by reasonable calibration and/or
statistical uncertainties.

It is also interesting to note that in currently
favored models (i.e., cosmic-concordance models with
$n\simeq1$), the optical depth to the surface of last scatter
cannot be too large, $\tau\lesssim0.2$.
A stochastic gravitational-wave background could mimic the
effect of reionization by supplying power on large angular
scales at which the CMB power spectrum is normalized to the COBE
amplitude.  Thus, the upper limits to $\tau$ can be translated
directly to upper limits to the gravitational-wave amplitude
${\cal T}$ (see \cite{KamKos99} for a precise definition) by
identifying $e^{-2\tau}$ with ${\cal S}/({\cal T}+{\cal S})$.
Doing so, the nominal limit $\tau\lesssim0.2$ suggests that no
more than one-third the large-angle power in the CMB is due to
gravitational-waves, and this improves slightly the limit to the 
gravitational-wave amplitude from COBE \cite{zibin}.

It has long been appreciated that the richness of the peak
structure in the CMB power spectrum will eventually allow
simultaneous determination of a number of cosmological
parameters \cite{jkks} when the CMB power spectrum is measured
with sufficient precision.  However, it has also been repeatedly
emphasized that a strong degeneracy in the
$(\Omega_m,\Omega_\Lambda,h)$ parameter space exists (e.g.,
\cite{ZalSelSpe97}), as indicated, for example, by the
elongation of the error ellipses forecast for MAP and Planck along the
$\Omega_m+\Omega_\Lambda$ line in the
$\Omega_m$-$\Omega_\Lambda$ parameter space (e.g., Fig. 2 in
Ref. \cite{OmegamLambda}).  In this
paper, we have noted that by implementing recent measurements
of the geometry, baryon density, and especially the Hubble
constant, we can break this degeneracy and thus link the
location of the second peak fairly robustly to the cosmological
constant.\footnote{The value of $h$ changes considerably as
one goes from one end of the aforementioned CMB ellipse in the
$\Omega_m$-$\Omega_\Lambda$ parameter space to the other end.}
This observation is additionally noteworthy given the
accumulation of independent evidence for some sort of dark
energy, the identification of the first peak, and the
approaching discovery of the second peak.  Thus,
by {\it visual inspection alone}, we may be able to learn
something significant about the cosmological constant once the
second peak is identified.  Of course, mapping the second peak
is also of the utmost importance as it will provide additional
confirmation of the paradigm of structure formation from
primordial adiabatic perturbations that underlies the entire
analysis.  We thus eagerly await the discovery of the second peak.

\medskip
We thank P. Ullio for useful comments.  We used CMBFAST
\cite{zs} to calculate the power spectra.  This work was
supported in part by the DoE, NSF, and NASA.  AB also
acknowledges the support of a Lee A. DuBridge Fellowship.

\end{document}